\documentclass{article}
\usepackage[utf8]{inputenc}
\usepackage{graphicx,pstricks}
\usepackage{amsmath}
\usepackage{amssymb}
\usepackage{graphics}
\usepackage{moreverb}
\usepackage{epsfig}
\usepackage{txfonts}
\usepackage{palatino}
\usepackage{setspace}
\usepackage{multirow}
\usepackage{csquotes}
\usepackage{geometry}
\usepackage{lscape}
\usepackage{booktabs}
\usepackage{rotating}
\usepackage{array}
\usepackage{natbib} 
\usepackage{longtable}
\usepackage{lipsum}
\usepackage{hangcaption}
\usepackage{enumitem}
\usepackage{textcomp}
\usepackage{gensymb}
\usepackage{caption}
\usepackage{subfig}
\usepackage{url}

\newcolumntype{P}[1]{>{\centering\arraybackslash}p{#1}}

\makeatletter
\pretocmd{\@sect}{\singlespacing}{}{}
\pretocmd{\@ssect}{\singlespacing}{}{}
\apptocmd{\@sect}{\doublespacing}{}{}
\apptocmd{\@ssect}{\doublespacing}{}{}
\makeatother

\renewcommand{\caption}[1]{\singlespacing\hangcaption{#1}\normalspacing}

\begin{document}

\begin{titlepage}

\title{Monetary Incentives, Landowner Preferences: Estimating Cross-Elasticities in Farmland Conversion to Renewable Energy
}

\author{ Chad Fiechter\footnote{Assistant Professor, Department of Agricultural Economics, Purdue University, West Lafayette}, Binayak Kunwar\footnote{PhD Student, Department of Agricultural Economics, Purdue University}, Guy Tchuente\footnote{Assistant Professor, Department of Agricultural Economics, Purdue University}}


\date{\today}

\maketitle

\begin{abstract}

This study examines the impact of monetary factors on the conversion of farmland to renewable energy generation, specifically solar and wind, in the context of expanding U.S. energy production. We propose a new econometric method that accounts for the diverse circumstances of landowners, including their unordered alternative land use options, non-monetary benefits from farming, and the influence of local regulations. We demonstrate that identifying the cross elasticity of landowners' farming income in relation to the conversion of farmland to renewable energy requires an understanding of their preferences. By utilizing county legislation that we assume to be shaped by land-use preferences, we estimate the cross-elasticities of farming income. Our findings indicate that monetary incentives may only influence landowners' decisions in areas with potential for future residential development, underscoring the importance of considering both preferences and regulatory contexts.
\end{abstract}

\bigskip

\textbf{\textit{Keywords}}: Instrumental variables; two-stage least squares;  multiple choices, agricultural landscapes; farm income; renewable energy; solar energy; .
\bigskip

\textbf{JEL}: C36, Q15, Q42.



 \end{titlepage}

 \section{Introduction}

The US is expanding renewable energy generated from solar and wind on farmland. From 2023 to 2025, solar and wind energy generation are expected to increase 75\%  and 11\%, respectively \citep{eia2024renewable1}. Some of the projected expansion will take place on farmland, removing the land from agricultural production, potentially increasing prices, and affecting the regional economy \citep{schultz2021renewable}. Additionally, scientists are concerned with efficiently locating, or siting, these renewable projects for neighboring ascetics \citep[e.g.][]{von2017measuring}, environmental implications \citep[e.g.][]{lubowski2006environmental, owen2004environmental}, and the preservation of food production \citep[e.g.][]{cunningham2024competition, zhuang2022land}. However, despite these potential externalities, lease contracts for solar and wind energy are established with heterogeneous private landowners.\footnote{In this paper we will use landowners and farmers interchangeably to refer to the individuals that could change the actual use of a farmland.} These externalities are currently addressed through heterogeneous county-level regulations. This study seeks to estimate the degree to which monetary factors affect the conversion of farmland to renewable energy generation, proposing a new econometric method that accounts for land use preference heterogeneity in landowners.

Landowners experience varying circumstances when considering a renewable energy lease, adding complexity to the estimation of the effects of changes in returns to farming. First, each landowners' alternative land use options vary. For example, landowners in counties with residential or commercial development potential may anticipate high long-term returns from this future development. In contrast, a landowner in a rural county may not consider the potential returns from residential or commercial development. Second, landowners can derive non-monetary utility from farming. Although solar and wind developments often offer farmland lease rates greater than the rates typically offered by tenant farmers \citep{cca2024agbarom}, the long-term utility of landowners for maintaining agriculture production may outweigh the monetary incentive. Third, the landowners' county may regulate the development of solar or wind projects or both. For instance, 15\% of US counties have stopped the development of solar and wind projects through regulation \citep{weise2024ban}. Lastly, solar development drastically alters the agricultural production system, often eliminating the potential for traditional row crops, whereas wind development converts a relatively small proportion of the land, preserving the potential for farming. The response of landowners to monetary factors is influenced by both observed factors, such as county regulations, and unobserved factors, like preferences. 

This study proposes a new econometric method to account for the heterogeneity in landowners' circumstances and causally identify the impact of short-term returns to farming on the decision to accept wind or solar leases. Specifically, we use the time-varying and cross-sectional variation in county-level wind and solar development regulations \citep{weise2024ban}, reported solar and wind installations, and the exogenous variation in growing degree days from \citet{schlenker2009nonlinear}. Our proposed empirical framework is inspired from the literature of multiple treatments and instrumental variables \citep{kirkeboen2016field, heckman2018unordered, bhuller20242sls}.

Our approach address the inherent endogeneity of returns to farming effects in landowners' land-use decisions and illustrate the importance of accounting for the actual land-use options. The estimation challenge stems not only from typical issues of correlated unobservables but also from the fact that landowners face a set of unordered alternatives. We demonstrate that having a valid instrument for farm return is insufficient to estimate the cross-return elasticity between options. Proper identification requires understanding how individuals rank these competing land-use options. Without this, the IV method cannot fully capture the cross-return elasticities or the effects of selecting one land use over another for any subgroup of the population.

Our aim is to accurately capture a policy-relevant estimate for the complex issue of land use regulation and demonstrate the importance of accounting for unordered alternative options. We identify the degree to which an increase in returns to farming impact the likelihood of land being converted to wind or solar energy production, or the cross-return elasticity of farming and solar or wind conversion. Although this study is the first to estimate this specific parameter, land use conversion is a classic economic problem\footnote{See \citep{meiyappan2014spatial, plantinga2021recent, zhao2020improving} commodity supply dynamics \citep{babcock2015extensive, brown2014merging, golub2012modeling, lin2019exploring,  marcos2017agricultural, miao2013impact, radeloff2012economic, swinton2011higher} and environmental impacts \citep{adjei2024nonlinear, bryan2013impact, claassen2011crop, closup2014wind, doidge2020role, huang2010econometric, plantinga2001supply, plantinga2002efficient, roberts2007enduring, rosenberg2024land, secchi2011land, stuart2013scaling}}. Related to our study, \citeauthor{lubowski2008drives}'s \citeyearpar{lubowski2008drives} findings demonstrate policymakers' difficulty in establishing land-use policy on farmland with monetary incentives. They find that federal conservation programs incentivize landowners to transition land out of agricultural production when returns to farming are low and then back to production when farming returns are high. ``By raising the profitability of cropland, the government increased acreage in crops and directly competed with itself in providing incentives for landowners to retire environmentally-sensitive cropland...'' \citep[pg. 545][]{lubowski2008drives}. 

We believe that our identified cross return elasticities for farming on solar and wind conversion provide practical information for land-use policymakers. First, landowners' willingness-to-accept solar and wind leases  only appear to be at the monetary decision margin for limited scenarios. The first scenario is for landowners who may anticipate the potential to develop their farms for residential or commercial purposes. Landowners in this scenario percieve the short-term returns to farming with the long-term option to develop the land similiar to the returns to solar or wind conversion. As mentioned earlier, lease rates for solar or wind energy are greater than typical returns to farming. As a result, landowners in the scenario either accept the solar and wind lease because the return is greater, or the landowner rejects the lease because they derive greater non-monetary utility from keeping the parcel in agriculture production. In either case, the landowners' decision is not at the monetary margin. We believe this scenario illustrates that monetary based policies aimed at guiding land-use decisions will be limited in effect. In the second scenario, landowners' willingness to accept wind and solar leases is directly related to the degree to which the conversion limits the ability to farm. As a result, wind and solar energy regulations are the main determinant of the potential expansion.

Additionally, we believe this paper makes contribution to the econometric literature, focusing on the methods of instrumental variables (IV) and treatment effects. We build on the work of \cite{imbens1994identification} who formally developed the LATE framework. They showed how instrumental variables could be used to estimate local treatment effects for compliers, a sub-population that adheres to the treatment assigned by the instrument. Our identification section also explores the use of the potential outcome framework in the presence of heterogeneous treatment effects and unordered choice options that could be dependent or independent variables. Thus, our paper follows the work of \cite{heckman2005structural, heckman2018unordered,blandhol2022tsls} who examined the conditions under which instruments are valid for identifying heterogeneous causal effects. We focus our attention on the context in which there are multiple options. Unlike in \citep{heckman2018unordered, lee2018identifying, kirkeboen2016field} where the unordered options are treatments, we are interested in the case where the unordered options are dependent variables and we are interested in the identification and estimation of the effect of an endogenous treatment on the option choices. We also consider the case where the options are independent variables. 

The remainder of the paper is organized as follows. Section \ref{sec:Id} presents our discussion of the identification of the policy-relevant treatment effect. Section \ref{sec:Back} describes the contracts and options offered to landowners when land-use changes are considered. In Section \ref{sec:Emp}, we present our empirical approach to the estimation of cross-income elasticity, while in Section \ref{sec:EstwR}, we present and discuss our estimated effects. Section \ref{sec:cls} presents the conclusions and recommendations derived from our analysis.

\section{Identification of the Farming Income Effects in the Presence of unordered Land Use Options}\label{sec:Id}

Our aim is to examine the degree to which an external increase in cash impact landowners' decision to convert farmland for different uses? We focus on the choice of land use for the primary outcomes of adopting wind or solar energy production to analyze how landowners respond to financial incentives. This section proposes an empirical methodology to estimate the effect of an exogenous income increase on the choice of unordered alternative land use. 

In estimating farm income elasticity of land-use, we must account for farm income being endogenous. As are result, we use an instrumental variable to estimate the income elasticity of land-use. Consider an instrumental variable $z_i$ (number of days with extreme heat, see \cite{bergman2020effect} and reference therein). 

Also, consider a sample of $N$ landowners. We use the dummy variables $d_i^A$ ,  $d_i^S$  $d_i^W$ and,  $d_i^O$ to represent the decision to use land for agriculture, solar panels, wind turbines, or other purposes (including residential) by landowner $i.$ 

We consider $Ic_i$ to be the farming income level of landowner $i$ , and $X_i$ a set of exogenous variables affecting farming income or the choice of a specific type of usage of the land.

A simple approach to estimate the farming income elasticity of land use involves using  the instrumental variable $z_i$ and estimating the following system for each potential use:
	$$d_i^k= \ohm^k+\omega^k  Ic_i+X_i\beta^k+\varepsilon_i^k ,  $$               $$ Ic_i=\theta + \pi z_i +X_i\rho +v_i ,$$ and
$k=A, S$, or $W$.\\
 We assume $X_i$ are exogenous and $z_i$ is a valid instrumental variable. Under classical assumptions, a 2SLS estimator of $\omega^k$ provides a consistent estimate of the income elasticity related to the choice of option $k$.  

Despite consistent estimation of $\omega^k$, it is difficult to give economic meaning to these instrumental variable (IV) estimates.  The challenge is that the option $k$ is compared to all other alternatives. We should note that, if some options are not available or the conversion of land from one use to option $k$ is largely easier than from another, it may not be policy relevant to estimate option $k$ versus all other options. To obtain policy-relevant income elasticities, identification of an economically relevant parameter does not only requires one instrument for income but also must address the issue that landowners choosing the same land use may have different available options (or next-best alternatives). It is, therefore, important to formalize the conditions under which the IV estimate can produce policy-relevant estimates.

\subsection{Instrumental Variable in Unordered Outcomes Choice Models}
 
This section explores the challenges related to the use of instrumental variable approach (IV) to estimate treatment effects when the outcomes are multiple and unordered. We provide guidance for our empirical section while showing the assumptions under which IV estimates can be given a causal interpretation as local average treatment effects (LATE) in settings with multiple unordered outcomes. For notational simplicity, we consider a case in which individuals choose between three alternatives. We will consider two main cases: first we assume that the option choice is exogenous in the determination of the outcome of interest, but the income elasticity is heterogeneous; second we assume that the outcome of interest is the option choice.  The potential outcome framework will be used to illustrate the causal nature of the IV estimator.

\subsubsection{Potential Outcomes: Model with Continuous Dependent variable}

We consider that there are three mutually exclusive outcomes \{0, 1, 2\} ( in our case 0$\simeq$ A, 1$\simeq$ S, 2$\simeq$W). For simplicity and to center our interest on the interpretation of IV (and OLS) , we suppress the individual index and also abstract any control variables. \\ 

The following  simple models could be postulated for each option:
 \begin{equation}
     y^k= \alpha_0^k+\delta^k  Ic+\varepsilon^k,
 \end{equation} with $k=0,1,$ or $2$.
 
 Let $y$ be the observed outcome: this scenario could be the use of the land or any other outcome. 
\begin{equation}
    y= y^0 + (y^1-y^0)d^1+ (y^2-y^0)d^2, 
\end{equation} 
where $d^k$ is a dummy variable taking the value $1$ if the option $k$ is observed and 0 otherwise.
 
\begin{equation}
    y= y_0 + (y_1-y_0)Ic
\end{equation}
For simplification,  assume that the returns to farming can take two values $Ic; ic\in \{0, 1 \}$ and, and as a result, $y_0$ and $y_1$ are the potential outcomes when the income level are respectively 0 and 1.

The corresponding structural equation of interest is: 
\begin{equation}\label{full_struc}
    y= \beta_0 + \beta_1 d^1 + \beta_2 d^2 +\theta_0  Ic+\theta_1 Ic \times d^1 + \theta_2 Ic \times d^2 + \psi,
\end{equation}
where $\psi$ is an $iid$ the error term and $\beta_0, \beta_1, \beta_2,\theta_0, \theta_1, \theta_2$ $\in R$  are unknown parameter of interest. 

 Under the assumption of exogeniety of $d^k$ and $Ic$, the parameters of the model in Equation (\ref{full_struc}) can be estimated using OLS.  
However, if $Ic$ is endogenously determined, additional assumptions are needed.  

The estimation by OLS of equation \ref{full_struc}  of the returns to farming elasticity of adopting  option 1  is $$E(y|  Ic=1, d^1=1 )- E(y| Ic=0, d^1=1)=E(y_1-y_0|Ic=1, d^1=1)+E(y_0|Ic=1, d^1=1)-E(y_0|Ic=0, d^1=1)$$.
$\Delta=y_1-y_0$ is the individual-level change the outcome $y$ resulting form a change in the level of income from 0 to 1. $E(y_1-y_0|Ic=1, d^1=1)$ is the expected cross return elasticity of adopting option 1.  

Our main objective is to correct for the selection bias that is shown in the above estimate if $E(y_0|Ic=1, d^1=1)-E(y_0|Ic=0, d^1=1) \neq 0$.

In the absence of selection bias, the OLS estimate of \(\theta_1\) captures \(E(\Delta|Ic=1, d^1=1)\), which denotes the change in outcomes for individuals who choose to adopt the same land use despite differing next-best alternatives. However, researchers typically observe only the chosen land use. Let \(d^{-j}\) denote the next best alternative of an individual, namely, the land use that would have been chosen if the option \(j\) had been removed from consideration.

The expression for \(E(\Delta|Ic=1, d^1=1)\) can be decomposed as:
\begin{align*}
   E(\Delta|Ic=1, d^1=1)&=&E(\Delta|Ic=1, d^1=1,d^{-1}=0 ) P[d^{-1}=0 |d^1=1]\\ &+&E(\Delta|Ic=1, d^1=1,d^{-1}=2 ) P[d^{-1}=2 |d^1=1] 
\end{align*}

It is challenging to give the OLS estimate of \(\theta_1\) a precise economic interpretation because it represents a weighted average of the changes in outcomes for individuals who choose the same land use option 1 but have different next-best alternatives. The average change in outcome across individuals with different next-best alternatives varies:
\[ E(\Delta|Ic=1, d^1=1, d^{-1}=0) \neq E(\Delta|Ic=1, d^1=1, d^{-1}=2) \]

This discrepancy arises because \(\Delta\) varies across individuals, and their ranking of land use options could partly depend on these idiosyncratic changes in the outcomes.

The economic interpretation of \(E(\Delta|Ic=1, d^1=1)\) becomes complex when everyone who selected option 1 would have chosen option 2 as their next-best alternative, resulting in \(P[d^{-1}=2 | d^1=1] = 1\). In such scenarios, \(E(\Delta|Ic=1, d^1=1)\) represents the average difference in outcomes between choosing option 1 versus options 0 and 2, while the relevant margin of choice is between options 2 and 1 only.

The OLS estimates of the farm income elasticity of land use  can vary either because of selection bias, differences in potential land use choice across income levels, or differences in weights across the next-best land use alternatives.

 \subsubsection{Potential Outcomes: The Model with the Choice of an Option as Dependent Variable}

For each option $k$, there are two relevant margins. For example, if option 1 is observed, it could be chosen by comparing it with option 0 or to option 2. Let $M_k$ a variable indicating the relevant margin being considered by the landowner when option $k$ is chosen.  

We assume that $M_k=1$ if the option $k$ is compared with the lowest numerical value of the remaining options.

$d^k=d^k(0)+ (d^k(1)-d^k(0))M_k$,
where $d^k(0)$ and $d^k(1)$ are respectively the potential outcomes of $d^k$ when the $M_k$ are respectively 0 and 1.
For each option, we also have
\begin{equation}
    d^k= d_0^k + (d^k_1-d^k_0)Ic,
\end{equation}
where $d^k_0$ and $d^k_1$  are respectively the potential outcomes of $d^k$ when the $Ic$ are 0 and 1.

We could then re-express $d^k$ in the potential outcome framework as follows:

$$d^k=d_0^k (0)+(d^k_0(1)-d^k_0(0))M_k+ ((d_1^k(0)-d^k_0(0))Ic+[(d_1^k(1)-d^k_0(1))Ic-(d_1^k(0)-d^k_0(0))Ic] M_k$$
or
$$d^k=d_0^k (0)+(d^k_0(1)-d^k_0(0))M_k+ ((d_1^k(0)-d^k_0(0))Ic+[(d_1^k(1)-d^k_0(1))-(d_1^k(0)-d^k_0(0))] Ic M_k$$.

For each of the choice option, the structural equation for the option choice model is:
\begin{equation}
    \label{eq:choi_str}
    d^k= \varrho_0^k + \varrho_1^k Ic + \varepsilon^k.
\end{equation}

The estimation by OLS of equation \ref{eq:choi_str}  of the cross return elasticity of farming and adopting  option $k$  is 
\begin{align}
    E(d^k|  Ic=1 )- E(d^k| Ic=0)&=&E(d^k_1-d^k_0|Ic=1)+E(d^k_0|Ic=1)-E(d^k_0|Ic=0)\\
    &=& E(d_1^k(0)-d^k_0(0)|Ic=1) +E(d^k_0|Ic=1)-E(d^k_0|Ic=0)\\
    &+& E[M_k (\{d_1^k(1)-d^k_0(1)\}-\{d_1^k(0)-d^k_0(0)\})|Ic=1].
\end{align}

Let $\Omega^k_0=d_1^k(0)-d^k_0(0)$ be the  individual-level cross-return elasticity of adopting option $k$ against option $0$.   The OLS estimation of $\varrho_1^k$ is unbaised  if : (a) $Ic$ is exogenous , and  (b) $E[M_k (\{d_1^k(1)-d^k_0(1)\}-\{d_1^k(0)-d^k_0(0)\})|Ic=1]=0$.\footnote{It can be noticed that if $\Omega^k_0=\Omega^k_1$ , which means that income elasticity does not depend on the margin being considered, the condition (b) holds. }

It is crucial to note that even in the absence of typical selection bias ($E(d^k_0|Ic=1)-E(d^k_0|Ic=0)=0$), if cross return elasticities are heterogeneous and depend on the margins being considered, the OLS estimator will produce biased estimates.

\subsection{Estimation of Farming Income Elasticity: What can IV Identity?}

This section examines the identification of the income elasticity in  the presence of a valid instrument. We show that identifying the option specific cross-return elasticity, in the case of multiple (more than 2) unordered options, requires additional assumptions than those needed in settings with binary choices.

We assume the existence of a binary IV, Z ( $z\in \{0,1\}$) for which the following assumptions hold. 

 \textbf{Assumption I1} (Exclusion) For all $k\in \{ 0,1,2\}$
$d^k_{ic,z}=d^k_{ic}$ and  $y^k_{ic,z}=y^k_{ic}$ for all  $ic,z$.\\
 
\textbf{Assumption I2} (Independence) For all $k\in \{ 0,1,2\}$
 ($d^k_{ic}, y^k_{ic}, Ic_z) \perp Z$ for all $ic,z$.\\
 
\textbf{Assumption I3} (Rank)  $E[z'ic]$ has full rank.\\

\textbf{Assumption I4} (Monotonicity)  $Ic_0\leq Ic_1$ 

These IV assumptions are similar to those found in the literature \citep{kirkeboen2016field,imbens1994identification}.

\subsubsection{IV Estimation with General Outcome }

The estimation using IV will be based on the following moment conditions:

$E[\psi z]=0$, $E[\psi zd^1]=0$, $E[\psi z d^2]=0$,  $E[\psi d^1]=0$, $E[\psi  d^2]=0$   and $E[\psi]=0$ for the estimation of the parameters in Equation (\ref{full_struc}).

From the moment conditions $E[\psi z]=0$, $E[\psi zd^1]=0$, $E[\psi z d^2]=0$,  $E[\psi d^1]=0$, $E[\psi  d^2]=0$   and $E[\psi]=0$ , using the potential outcomes framework and the independence assumption, we derive a set of equations to be solved to obtain the parameters of  Equation (\ref{full_struc}) . 

Note that \begin{equation}
    y= y^0 + (y^1-y^0)d^1+ (y^2-y^0)d^2, 
\end{equation} 
where $d^k$ is a dummy variable taking the value $1$ if the option $k$ is observed and 0 otherwise.
 
\begin{equation}
    y^k= y^k_0 + (y^k_1-y^k_0)Ic,
\end{equation}
for $k\in \{ 0,1, 2\}$,  and  $Ic= Ic_0+ (Ic_1-Ic_0)z.$

Let $\Delta^0=y_1^0-y_0^0$ be the individual level of elasticity under option 0,  $\Delta^1=(y_1^1-y_0^1)-(y_1^0-y_0^0)$ the change in the individual elasticity resulting from moving from option 0 to 1,   $\Delta^2=(y_1^2-y_0^2)-(y_1^0-y_0^0)$ the change in the individual elasticity due to a switch  from option 0 to 2. Consider $\rho_1=(y_0^1-y_0^0)$ to be the return to choosing option 1 versus option 0 when $Ic=0$  and,  $\rho_2=(y_0^2-y_0^0)$  be the return to choosing option 2 versus option 0 when $Ic=0$. 

\begin{align}  \nonumber
    \psi&=& (y_0^0-\beta_0) +  [\rho_1-\beta_1] d^1 +[\rho_2-\beta_2] d^2 +[\Delta^0 -\theta_0]Ic +[\Delta^1-\theta_1] Ic\times d^1+ [\Delta^2-\theta_2] Ic\times d^2\\  \nonumber
   & =& (y_0^0-\beta_0)  + [\rho_1 -\beta_1]d^1 +[\rho_2-\beta_2] d^2 +[\Delta^0-\theta_0] [Ic_0+ (Ic_1-Ic_0)z]\\\nonumber &+&[\Delta^1-\theta_1] d^1\times [Ic_0+ (Ic_1-Ic_0)z]+ [\Delta^2-\theta_2]d^2\times [Ic_0+ (Ic_1-Ic_0)z]
\end{align}
    
The moment conditions couple with the independence assumption to imply 
\begin{equation} \label{Eq_sol1}
  E\left[ \{(\Delta^0-\theta_0)  + (\Delta^1-\theta_1) d^1 +(\Delta^2-\theta_2)d^2 \} (Ic_1-Ic_0)  \right]  =0
\end{equation}
\begin{equation} \label{Eq_sol2}
  E\left[ \{(\Delta^0-\theta_0) + (\Delta^1-\theta_1)d^1   \} (Ic_1-Ic_0)d^1  \right]  =0
\end{equation}
\begin{equation} \label{Eq_sol3}
  E\left[ \{(\Delta^0-\theta_0) + (\Delta^2-\theta_2)d^2   \} (Ic_1-Ic_0) d^2 \right]  =0
\end{equation}

\textbf{Proposition 1.} Under assumptions I1 to I4, solving the system of equations (\ref{Eq_sol1}) to (\ref{Eq_sol3}) to obtain the cross return elasticity $\theta_0, \theta_1,$ and $\theta_2$, we can show that $\theta_j$, $j=0,1,2$ is a linear combination of $\Delta^0$, $\Delta^1$ and $\Delta^2$. In fact, $\theta_0= E\left[ \Delta^0     |(Ic_1-Ic_0)=1, d^1=d^2=0\right]$, this is a LATE on the set of compliers who have chosen option 0. 
$$\theta_1= E\left[ \{(\Delta^0-\theta_0)  + \Delta^1   \}   |(Ic_1-Ic_0)=1, d^1=1\right], $$ 
and
$$\theta_2= E\left[ \{(\Delta^0-\theta_0)  + \Delta^2   \}   |(Ic_1-Ic_0)=1, d^2=1\right]. $$  
\textbf{Proof of Proposition 1:}
\begin{align} \nonumber
 0&=&  E\left[ \{(\Delta^0-\theta_0)  + (\Delta^1-\theta_1) d^1 +(\Delta^2-\theta_2)d^2 \} (Ic_1-Ic_0)  \right]  \\ \nonumber
  &=& E\left[ \{(\Delta^0-\theta_0)  + (\Delta^1-\theta_1) d^1 +(\Delta^2-\theta_2)d^2 \}   |(Ic_1-Ic_0)=1\right] \\ \nonumber
  &=& E\left[ \{(\Delta^0-\theta_0)  + (\Delta^1-\theta_1)   \}   |(Ic_1-Ic_0)=1, d^1=1\right] Pr(d^1=1)\\ \nonumber
  &+&E\left[ \{(\Delta^0-\theta_0)   +(\Delta^2-\theta_2) \}   |(Ic_1-Ic_0)=1, d^2=1\right] Pr(d^2=1)\\ \nonumber
  &+& E\left[ \{(\Delta^0-\theta_0)  \}   |(Ic_1-Ic_0)=1, d^1=d^2=0\right] Pr(d^1=d^2=0)
\end{align}

Similarly we have 
\begin{align} \nonumber
 0&=&  E\left[ \{(\Delta^0-\theta_0) + (\Delta^1-\theta_1)d^1   \} (Ic_1-Ic_0)d^1  \right]  \\ \nonumber
  &=& E\left[ \{(\Delta^0-\theta_0) + (\Delta^1-\theta_1)d^1   \}d^1| (Ic_1-Ic_0)=1  \right]\\ \nonumber
  &=& E\left[ \{(\Delta^0-\theta_0)  + (\Delta^1-\theta_1)   \}   |(Ic_1-Ic_0)=1, d^1=1\right] \\ \nonumber
\end{align} 
and
\begin{align} \nonumber
 0&=&  E\left[ \{(\Delta^0-\theta_0) + (\Delta^2-\theta_2)d^2   \} (Ic_1-Ic_0)d^2  \right]  \\ \nonumber
  &=& E\left[ \{(\Delta^0-\theta_0) + (\Delta^2-\theta_1)d^2   \}d^2| (Ic_1-Ic_0)=1  \right]\\ \nonumber
  &=& E\left[ \{(\Delta^0-\theta_0)  + (\Delta^2-\theta_2)   \}   |(Ic_1-Ic_0)=1, d^2=1\right]. \\ \nonumber
\end{align}

Thus,  $E\left[ \{(\Delta^0-\theta_0)  \}   |(Ic_1-Ic_0)=1, d^1=d^2=0\right] Pr(d^1=d^2=0)=0$, and  $\theta_0= E\left[ \Delta^0     |(Ic_1-Ic_0)=1, d^1=d^2=0\right]$, this result is a LATE for a complier who has chosen option 0. 

From $E\left[ \{(\Delta^0-\theta_0)  + (\Delta^1-\theta_1)   \}   |(Ic_1-Ic_0)=1, d^1=1\right]=0$, we can obtain $$\theta_1= E\left[ \{(\Delta^0-\theta_0)  + \Delta^1   \}   |(Ic_1-Ic_0)=1, d^1=1\right]. $$ 
Similarly, we can show that 
$$\theta_2= E\left[ \{(\Delta^0-\theta_0)  + \Delta^2   \}   |(Ic_1-Ic_0)=1, d^2=1\right]. $$ This ends the proof of Proposition 1.

Proposition 1 shows that the IV estimation of the parameters in equation (\ref{full_struc}) can only identify the cross-return elasticity of one option ($\theta_0$) for a group in the population. The cross-return elasticity of other options is not identified by any individual or group. For example, if you are comparing three land uses (agricultural, wind or solar) and  $y$ is a general outcome of interest, based on Proposition 1, IV estimation will not be able to tell us if the cross-return elasticity is larger for those who choose agriculture versus wind turbines or solar panel instead of production agriculture. The IV estimation identifies the weighted average individual elasticities.

\textbf{Proposition 2.} Under assumptions I1 to I4, solving the system of equations (\ref{Eq_sol1}) to (\ref{Eq_sol3}) to obtain the income elasticity $\theta_0, \theta_1,$ and $\theta_2$.
(i) If $\Delta_0$ is homogeneous for all individuals,

(ii) or  if 
\begin{align}
    \theta_0&=&E\left[ \Delta^0     |(Ic_1-Ic_0)=1, d^1=d^2=0\right]\\ &=&E\left[ \Delta^0    |(Ic_1-Ic_0)=1, d^1=1\right]\\&=& E\left[ \Delta^0    |(Ic_1-Ic_0)=1, d^2=1\right],
\end{align}

Then, $$\theta_1= E\left[   \Delta^1     |(Ic_1-Ic_0)=1, d^1=1\right], $$ 
and
$$\theta_2= E\left[  \Delta^2      |(Ic_1-Ic_0)=1, d^2=1\right]. $$  

Proposition 2  (i)  shows that if the elasticities, in one option,  are common for all individuals, the IV estimation of Equation (\ref{full_struc}) identifies the elasticities.  In the result (ii), the identification of $\theta_1$ and $\theta_2$ rely the homogeneity of LATE of income, under option 0,  for the complier who chooses option 0, 1 or 2.  
Although common elasticity could be reasonable in some cases, there is much empirical evidence of heterogeneous effects \citep{gebremariam2018heterogeneous,kirkeboen2016field,djebbari2008heterogeneous}  and individuals often make decisions based on these heterogeneous effects.  

Given the exogenous nature of the option choice, as assumed above, the main challenge for identification is the heterogeneity of treatment effects.

\subsubsection{IV Estimation in the Model for the Choice of Options }\label{sec:idenres}

The estimation of the parameter in equation (\ref{eq:choi_str}) will use the moments. $E[\varepsilon^k z]=0$ and $E[\varepsilon^k]=0$ for all $k$, 
$$d^k=d_0^k (0)+(d^k_0(1)-d^k_0(0))M_k+ ((d_1^k(0)-d^k_0(0))Ic+[(d_1^k(1)-d^k_0(1))-(d_1^k(0)-d^k_0(0))] Ic M_k$$,
and  $Ic= Ic_0+ (Ic_1-Ic_0)z.$

Let $\Omega^k_0=d_1^k(0)-d^k_0(0)$ be the  individual-level cross return elasticity of adopting option $k$ against option $0$,  $\Gamma_0^k=d^k_0(1)-d^k_0(0)$ is the individual preference between the option $k$ and the lowest numerical value of the remaining options when the income is at 0. 

The structural equation for the option choice model is:
\begin{equation}
    d^k= \varrho_0^k + \varrho_1^k Ic + \varepsilon^k.
\end{equation}
Thus we have 
\begin{align} 
    \varepsilon^k&=& (d_0^k(0)-\varrho_0^k )+ (\Omega_0^k-\varrho_1^k )Ic + \Gamma_0^k M_k+ [\Omega_1^k-\Omega_0^k] M_k\times Ic\\ 
    &=&(d_0^k(0)-\varrho_0^k )+ (\Omega_0^k-\varrho_1^k ) [Ic_0+ (Ic_1-Ic_0)Z]+ \Gamma_0^k M_k+ [\Omega_1^k-\Omega_0^k] M_k\times [Ic_0+ (Ic_1-Ic_0)z]
\end{align}
The moment condition $E[\varepsilon^k z]=0$ for all $k$ gives:
\begin{equation} \label{Eq_Choi}
  E\left[ \{(\Omega_0^k-\varrho_1^k )  + [\Omega_1^k-\Omega_0^k] M_k \} (Ic_1-Ic_0)  \right]  =0.
\end{equation}

Thus,  $\varrho_1^k =E[\Omega_0^k  + (\Omega_1^k-\Omega_0^k) M_k|Ic_1-Ic_0=1 ]/Pr(Ic_1-Ic_0=1)$.

\textbf{Proposition 3.} Under assumptions I1 to I4, for each option $k$, solving the system of equations (\ref{Eq_Choi}) to obtain the criss return elasticity $\varrho_1^k =E[\Omega_0^k  + (\Omega_1^k-\Omega_0^k) M_k|Ic_1-Ic_0=1 ]$. 

(i) If for all $k$ the cross return elasticities are homogeneous ( $d^k_1-d^k_0$, are constant for all individuals), then the IV estimator identifies the LATE.

(ii) If $\Omega_1^k=\Omega_0^k$ , the option you are leaving is irrelevant in determining the cross-return elasticity of the choice, then $\varrho_1^k =E[\Omega_0^k |Ic_1-Ic_0=1 ]$ identifies the LATE.

(iii) if $M_k=0$ (restrictive preferences) then $\varrho_1^k =E[\Omega_0^k |Ic_1-Ic_0=1 ]$ identifies the LATE.

 Proposition 3 highlights the importance of accounting for the margin being considered when unordered multiple alternatives are present. We have show that the IV estimator needs additional assumptions to deliver an economically meaningful estimand. 

 Result (i) and (ii) discuss the importance of homogeneity of treatment effects. And, result (iii) shows that if we impose an order of preference to the remaining options, the IV estimator of Equation (\ref{eq:choi_str}) identifies the LATE. 

 In practice, if we have a ranking of options (preference) for each individual we could use the next-best alternatives with  assumptions I1 and I4 to identify the LATE. When we restrict the sample to individuals with the same next-best, those who are induced to change income by a change in the instrument come from a particular option. This result means that we can assign a value to $M_k$.

 The identification result in Proposition 3 (iii) will be used in our empirical analysis.  Our design has two main parts: (i) use an instrument to correct for selection bias, and (ii) construct a ranking of options and use the next-best alternative to select the appropriate margin.

\section{Background: Land uses and Land Conversion Contracts }\label{sec:Back}

\begin{figure}
    \centering
    \subfloat[Wind]{\includegraphics[width=6.5cm]{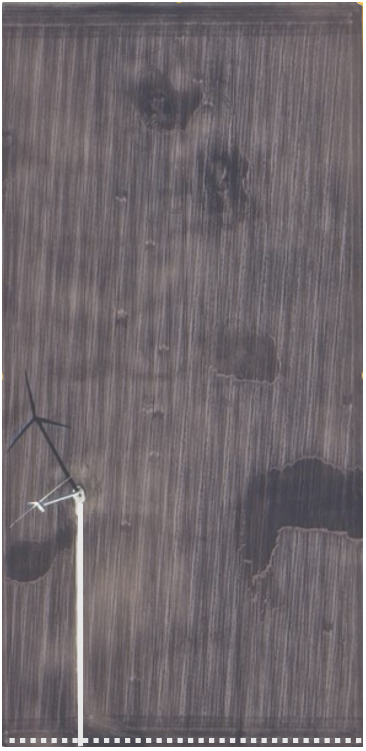} }
    \qquad
    \subfloat[Solar]{\includegraphics[width=7.1cm]{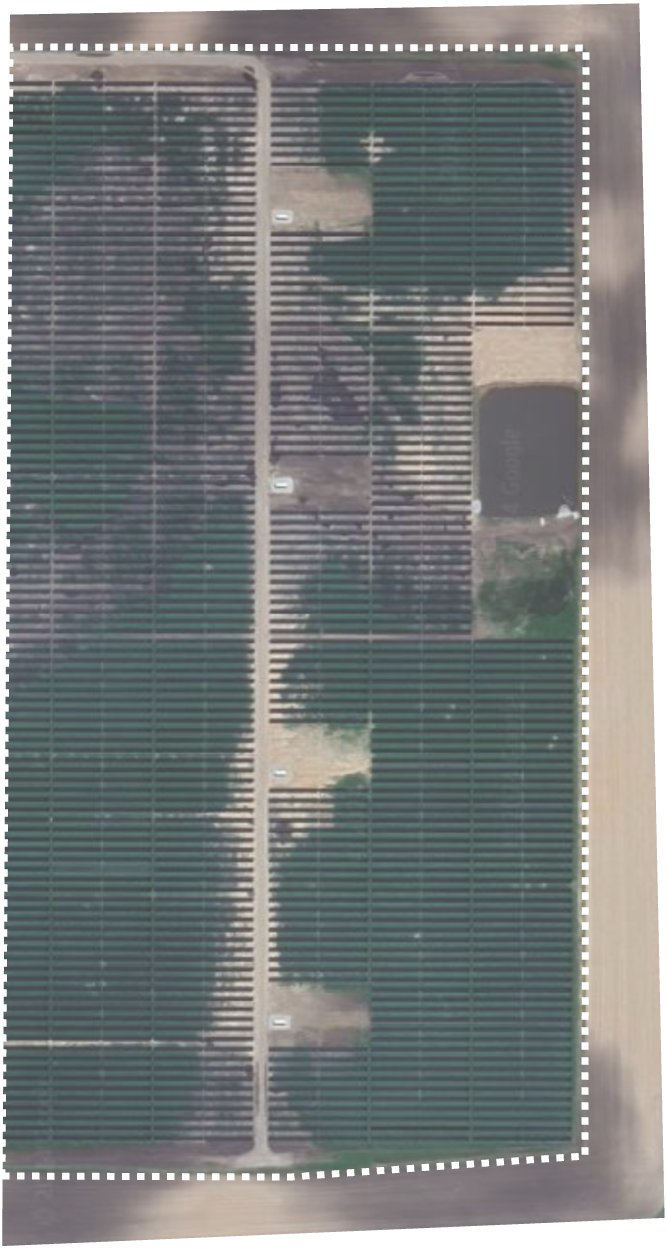} }
    \caption{A representative 80 acre farm with wind or solar energy generation (\textit{images not to scale})}
    \label{lease_sheme}
\end{figure}

For the sake of illustration, figure \ref{lease_sheme} represents an 80 acre Indiana farm with wind or solar energy generation. This study assumes that there are three potential uses of farmland: farming, wind energy generation, and solar energy generation. The following section details the factors within typical leasing arrangements as provided by experts in the renewable energy generation industry. Common annual cash lease agreements for crop production range from an annual payment of \$250 to \$350 per acre in Indiana, and this value serves as the residual option for farmland owners. The details of each of these leases are discussed below.

We compare two types of wind lease and their representative annual return. Initially, a landowner is offered an option contract during the development phase of a project. The option has a typical duration of five to seven years and results in an annual payment of \$5 to \$50 per acre. In addition, these projects are often contingent on the aggregation of large amounts of farmland (more than 5,000 acres), and key landowners may be offered one-time signing bonuses of \$1,000 to \$10,000. At the end of the option period, if the project has not entered the construction phase, the lease terminates. If the project proceeds to the contruction phase, a new payment scheme is triggered. There is substantial difference in the payment scheme during the construction phase (typically less than two years) ranging from the original option payment (\$10 to \$50 per acre) to full operation payment. Despite substantial variation, it is more typical that a landowner receives a payment valued between their option and operation payment. The construction payment is intended to offset the lost opportunity of leasing the farm for agricultural production.

The operation phase of wind energy generation leases defines the difference between the two payment schemes. The first payment scheme is limited to properties with the physical presence of a wind turbine, whereas the second scheme compensates each aggregated acre of the entire project. In the first scheme, landowners are paid according to the megawatts generated by the constructed turbine. For example, current turbines generate 4 to 5 megawatts annually. Payments range from \$5000 to \$10,000 per megawatt or \$20,000 to \$50,000 per turbine. Under this scheme, there are no additional per acre payments for the land without a turbine. In the second scheme, the presence of a turbine results in an annual payment of \$1,0000 to \$3,000 and an additional payment per acre on the entire property, similar to the development period (\$10 to \$50). The second scheme allows landowners whose property does not include a wind turbine to profit from energy generation. In the first scheme, these landowners are not compensated. This feature is helpful in securing the needed acres for development. 

In both schemes, the landowner is also paid for access roads and underground power transmission lines. Payments for access roads and power transmission are in the range of \$1 to \$2 per linear foot. The operations phase is typically 25 to 35 years with an option to extend for 10 to 20 additional years. Other key features of wind power generation are annual payment escalation, decommissioning security, and compensation for property tax increases. The annual payment escalators are fixed at one to two percent or variable and set according to the Consumer Price Index. Often, the variable escalators are capped at five or six percent.  

Solar energy generation leases exhibit the same structure of option phase, construction, and operation phase, but differ in payment schemes. Similar to the option phase for wind, the solar lease options specify a per acre annual payment of \$10 to \$50 per acre. Although solar developments typically requires less aggregated farmland than wind projects, key landowners may receive a one-time signing bonus in the range of \$1,000 to \$10,000. Payments for the construction phase follow a similar structure, and landowners receive payment valued in the range of their option to full operation payment. In the operation phase, landowners receive annual per acre payments in the range of \$1,000 to \$1,700 for the land they are no longer able to lease for crop production. Some leases also include a \$20 to \$40 payment for land that is not used in solar development, but included in the initial option. The operation phase duration is often 25 to 35 years, with the option to extend for 10 to 20 additional years. Lastly, annual payment escalators, decommissioning securities, and compensation for property tax increases are included in the solar lease.

Table \ref{lease_spec} provides and illustration of the annual lease payments for crop production, the operation phase of wind power generation, and the operation phase of solar power generation for the farmland parcels specified in figure \ref{lease_sheme}. For the calculations of table \ref{lease_spec}, we chose the midpoint of each payment range described previously. Additionally, these calculations are for the operations phase of wind and solar power generation, as implied by the options phase of the corresponding leases, there is significant probability that a lease does not progress to construction and operation. For landowners, they must consider the possibility that their land may not reach operation, and if it does, a wind turbine may not be constructed on their land or the amount of acreage used for solar panels may be minimal. Given these uncertainties, each parcel may generate \$24,000 in the residual option, crop production lease. Under the two wind power generation schemes, the landowner will recieve more income, with \textit{scheme 1} being a more risky outcome than \textit{scheme 2}. The premium represented in \textit{scheme 1} is contingent on a wind turbine being constructed on the property. Solar power generation in our specific example represents more than 75\% of the parcel being used for solar development. This feature is uncertain and likely depicts a favorable outcome. Often less farmland is used. However, the payment for solar energy generation in our illustration is greater than in other scenarios. 

\begin{table} [!htbp]
\centering
  \caption{Representative Lease Specifications for Crop Production, Wind Power Generation, and Solar Power Generation on Indiana farmland in 2024}
  \label{lease_spec}
\begin{tabular}{@{\extracolsep{5pt}}lcccc} 
\\[-1.8ex]\hline 
\hline \\[-1.8ex] 
            \\[-1.8ex]\hline 
            Lease & Quantity & Payment & Subtotal & Total\\  
            \\[-1.8ex]\hline 
    Crop Production & \multicolumn{1}{l}{80 acres} & \$300 &  & \$24,000 \\ 
    \addlinespace
    Wind Power &   &  &  &  \\ 
    \multicolumn{1}{r}{\textit{scheme 1}} &  \multicolumn{1}{l}{5 MW - 1 turbine} & \$7,500 & \$37,500 &  \\ 
     &  \multicolumn{1}{l}{1,600 ft - access road} & \$1.50 & \$2,400 &  \\ 
     & \multicolumn{1}{l}{1,300 ft - power transmission} & \$1.50 & \$1,950 &  \\ 
     & \multicolumn{1}{l}{77 acres - crop production lease} & \$300 & \$23,100 &  \\
     &   &  &  & \$64,950 \\
     \addlinespace
    \multicolumn{1}{r}{\textit{scheme 2}} &  \multicolumn{1}{l}{5 MW - 1 turbine} & \$2,000 & \$2,000 &  \\ 
     &  \multicolumn{1}{l}{1,600 ft - access road} & \$1.50 & \$2,400 &  \\ 
     & \multicolumn{1}{l}{1,300 ft - power transmission} & \$1.50 & \$1,950 &  \\ 
     & \multicolumn{1}{l}{80 acres - payment} & \$30 & \$2,400 &  \\ 
     & \multicolumn{1}{l}{77 acres - crop production lease} & \$300 & \$23,100 &  \\
     &   &  &  & \$31,850 \\
     \addlinespace
      Solar Power &   &  &  &  \\ 
     & \multicolumn{1}{l}{61 acres - solar power generation} & \$1,250 & \$76,250 &  \\ 
     & \multicolumn{1}{l}{19 acres - crop production lease} & \$300 & \$5,700 &  \\
      
    &   &  &  & \$81,950 \\
            \hline
            \multicolumn{5}{l}{\footnotesize Source: Author conversation with Indiana wind and solar power generation developers}\\
            \end{tabular}        
\end{table}

Solar and wind energy generation differ in the key dimension of amount of land fully converted to energy production. Using current technology, every acre with solar panels is not usable in crop production. This is not a characteristic of wind energy generation. A turbine typically uses less than one or two acres.  Although turbines and access roads add challenge to crop production, they occupy minimal farmland. In discussions with solar and wind developers, this key dimension is important for understanding landowner decisions. Anecdotes suggest that solar energy generation and the full conversion of farmland away from crop production are viewed as antithetical to the ideology of many landowners. These landowners generate a significant non-monetary benefit from using their land in crop production.  

Lastly, the payments described in table \ref{lease_spec} neglect the dynamic interactions with lease duration. Although crop production leases are often multi-year \citep{usda2014total}, a landowner will experience the ability to renegotiate terms frequently. From 2023 to 2024 crop production leases for top quality farmland in Indiana have increased by 2.3 percent \citep{kuethe2024purdue}. Wind and solar leases are 25 to 35 years contracts with options to extend 10 or 20 additional years. As a result, a landowner who accepts a farmland lease may never have an opportunity to renegotiate. Anecdotes suggest that the long duration of wind and solar leases may be attractive to absentee landowners who want to bequest farmland assets to their family. In contrast, those landowners who experience more utility from their connection with production agriculture likely view the longer duration as detrimental. 

\section{Empirical Approach} \label{sec:Emp}

We estimate the cross return elasticity with respect to land conversion to wind and solar energy using wind and solar energy facilities data obtained from United States Geological Survey (USGS) and Lawrence Berkeley National Laboratory(LBNL). USGS, in collaboration with LBNL provides U.S. Large-Scale Solar Photovoltaic Database (USPVDB) and U.S. Wind Turbine Database (USWTDB). These datasets contain information on name, year of installation, location, area and capacity of solar facilities and wind turbines, respectively. We collect solar and wind data for the ``corn belt'' in the U.S. heartland region, including county-level data from Illinois, Indiana, Iowa, Missouri, and Ohio. This dataset covers the period from 2015 through 2022. Additionally, we obtain county-level annual corn yield data, measured in bushels per acre harvested, from the National Agricultural Statistics Service (NASS) of the United States Department of Agriculture (USDA).  

Our main empirical analysis employs an instrumental variable approach to relate exogenous weather-driven changes in corn yield to adoption of solar and wind facilities on farmland. Based on prior literature that demonstrates corn's sensitivity to temperature variations during the growing season (April through September), we use number of days within this period where average daily temperature exceeds 83 degree Fahrenheit (83\degree F). A similar approach has been used by \citet{bergman2020effect} to study the effect of cash flow shocks during financial crisis. In our case, the cross sectional variation in days with high temperature directly impact returns to corn farming. These short-term returns to farming corn are indirectly related to the decision to accept a wind or solar lease. However, we do not believe that the cross sectional variation in days with high temperature directly impacts landowners' decision to accept a wind or solar lease. As a result, we believe this variable satisfies the exclusion restriction.

Specifically, for our analysis, average daily temperature for each county is collected from National Oceanic and Atmospheric Administration (NOAA) for the period of 2015 to 2022. NOAA obtain the data from weather stations, that are then weighted by area to compute county-wide values. We then calculate the county-level number of days with temperature above 83\degree F during the corn growing season.

In addition to income from corn yield, our model includes several control variables at the county level, including Gross Domestic Product (GDP), labor force statistics, unemployment rate, and median age. GDP data for each state is obtained from U.S. Bureau of Economic Analysis (BEA), while labor force and unemployment data is obtained from U.S. Bureau of Labor Statistics (BLS). Population data is collected from the U.S. Census Bureau. 

\begin{figure}[htbp]
\begin{centering}
\caption{Study area}
\includegraphics[scale=.62]{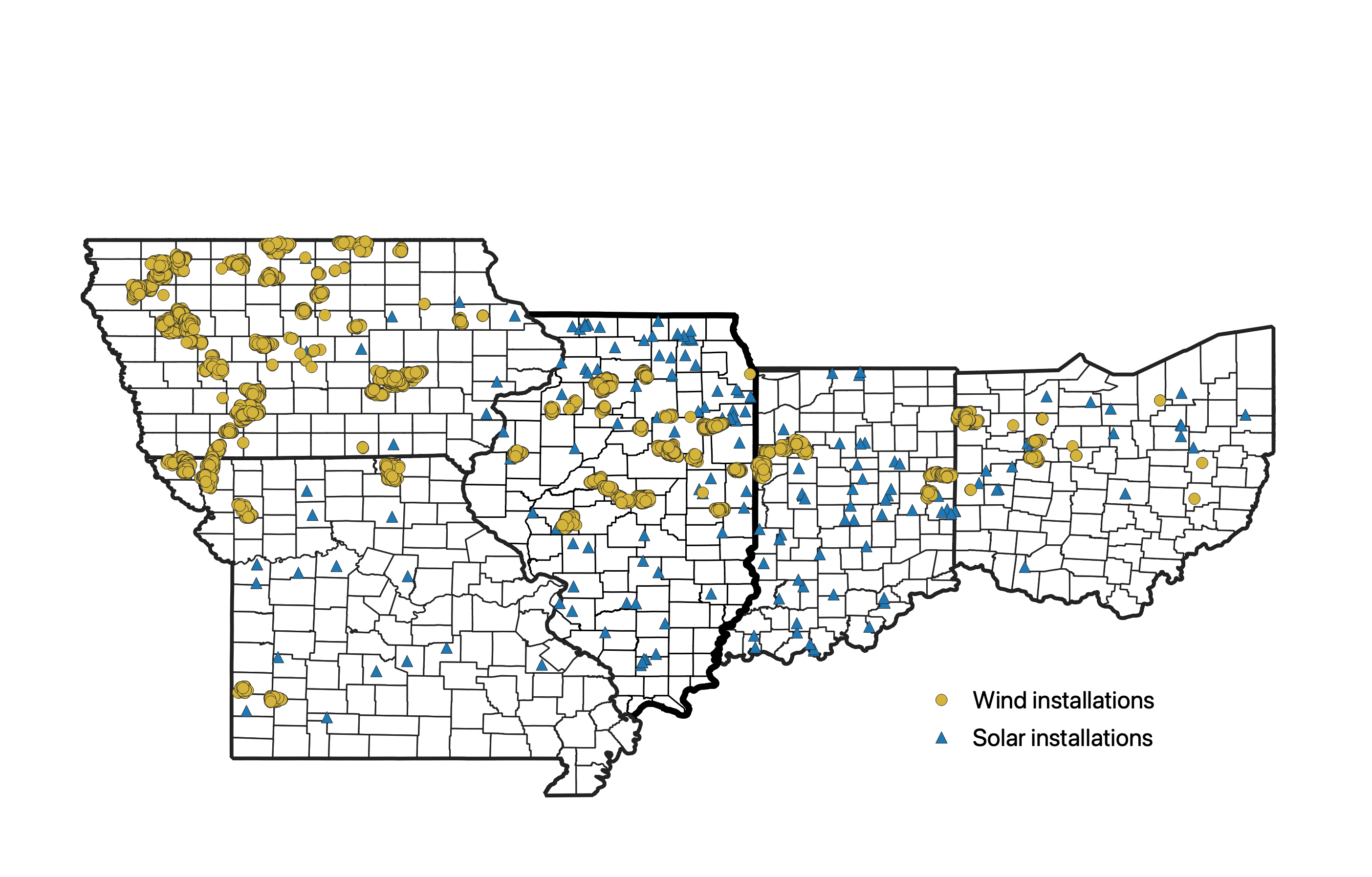}
\label{study area}
\end{centering}
\end{figure}

Table \ref{summarystat} reports the summary statistics of the economic variables used in our model. The variables SLR and WND denote the dummy variable for solar and wind adoption indicating adoption of solar and wind energy, where 0 represents no adoption and 1 represents adoption. In our study, the mean adoption for solar and wind energies are 0.13 and 0.11, respectively. The average number of growing season days with an average temperature exceeding 83 \degree F is 2.87, with a standard deviation of 4.70. Additionally, the average corn yield in our study area is 175.29 bushels per acre, with a standard deviation of 27 bushels per acre harvested. The averages of the natural logarithm of corn income and county GDP are 6.6 and 14.09, respectively. In addition, the mean values for the county population, median age, total labor force and unemployment rate are 81,155.38, 41.41 years, 40,888.49 individuals and 4. 48\%, respectively. Finally, the mean capacity of power transmission stations and the average number of power transmission stations per county are 2,499 megawatts and 12.08, respectively. The study area is shown in Figure \ref{study area}.

\begin{table} [!htbp]
\centering
  \caption{Summary Statistics}
  \label{summarystat}
\begin{tabular}{@{\extracolsep{5pt}}lcccc} 
\\[-1.8ex]\hline 
\hline \\[-1.8ex] 
            \\[-1.8ex]\hline 
            Variable & Mean & SD & Min & Max\\  
            \\[-1.8ex]\hline 
    SLR &  0.13 & 0.34 & 0 & 1 \\ 
    WND &  0.11 & 0.31 & 0 & 1 \\ 
    Days above 83 &  2.87 & 4.70 & 0 & 40 \\ 
    Corn yield &  175.29 & 26.95 & 56.10 & 246.70 \\ 
    log(Income) & 6.60 & 0.31 & 5.52 & 7.34 \\ 
    log(GDP) & 14.09 & 1.35 & 10.79 & 20 \\
    Population &  81155.38 & 271784.64 & 1955 & 5261249 \\ 
    Median age &   41.41 & 3.75 & 25.60 & 56.30 \\ 
    Labor force &  40888.49 & 139451.67 & 1138.00 & 2689206.00 \\ 
    Unemployment rate &  4.48 & 1.62 & 1.60 & 12.80 \\ 
            \hline
\multicolumn{5}{l}{\textbf{Source:} The sources of data utilized in our study are shown in Table \ref{data_source}.} \\
            \end{tabular}
        
\end{table}
\begin{table} [!htbp]
\centering
  \caption{Data Sources}
  \label{data_source}
\begin{tabular}{@{\extracolsep{5pt}}lc} 
\\[-1.8ex]\hline 
\hline \\[-1.8ex] 
            \\[-1.8ex]\hline 
            Dataset & Sources\\  
            \\[-1.8ex]\hline 
    U.S. Solar Photovoltaic Database &  United States Geological Survey (USGS) \\ 
    U.S. Wind Turbine Database &  United States Geological Survey (USGS)\\ 
    Corn yield &  USDA's National Agricultural Statistics Service (NASS)  \\ 
    Weather shocks & National Oceanic and Atmospheric Administration (NOAA) \\ 
    GDP & U.S. Bureau of Economic Analysis (BEA) \\ 
    Population &  U.S. Census Bureau \\ 
    Labor force & U.S. Bureau of Labor Statistics (BLS) \\
    Unemployment rate & U.S. Bureau of Labor Statistics (BLS) \\
    County Metro status & USDA's Economic Research Service (ERS) \\
    County Renewable Energy Regulations & USAtoday \\
            \hline
            \end{tabular}
\end{table}


\section{Estimation of Cross Return Elasticity: IV for farm returns and options ranking} \label{sec:EstwR}

In this section, we propose the estimation of cross return elasticity using insight of Proposition 3. First, we present a ranking of potential land use option base on urban versus rural nexus of  counties and the legislation on development of solar and wind energy production. This ranking implies the next-best alternative for landowners' in each county. Second, we estimate equation (\ref{eq:choi_str}), conditional on the implied common next-best alternative. 

\subsection{Land Use Option Ranking}
For ranking land use preference, we consider urban and rural status of counties in combination with renewable energy development regulations. To capture the urban and rural status, we use county-level typology codes provided from the USDA's Economic Research Service (ERS). The typology code captures a range of economic and social characteristics. Similarly, wind and solar regulation data are sourced from the USAToday Green Energy Nationwide Bans article \footnote{https://www.usatoday.com/story/news/investigations/2024/02/04/green-energy-nationwide-bans/71841275007/}. This article categorizes solar and wind energy regulations into four distinct groups: \textit{Ban}, \textit{Difficult to Permit}, \textit{Moratorium}, and \textit{Impediments}. The category \textit{Difficult to Permit} indicates counties where renewable energy projects are not banned, but face extreme regulations that deter their development. A \textit{Moratorium} is used by counties to buy time to draft new zoning and regulations for wind and solar development, sometimes to establish reasonable laws to permit renewable energy and sometimes to craft bans. \textit{Impediments} are specific restrictions that hinder renewable energy development. For instance, some counties allow turbines but require significant distances from property lines. In 2014, Connecticut mandated setbacks of at least 2.5 times a turbine’s height (approximately 1,250 feet), resulting in no new wind projects. In Indiana, voluntary renewable energy zoning regulations include a maximum sound limit of 50 decibels, comparable to the noise level of a household refrigerator. Also, resulting in no new projects.

For our ranking analysis, we assign codes 2, 3, 4, and 5 for Impediments, Moratorium, Difficult to Permit, and Ban, respectively, with 1 indicating no such restrictions.

In rural counties, agricultural land use is ranked first. We then assess renewable energy regulations to rank solar energy, wind energy, and residential land use. If either solar or wind energy is banned, the banned category is ranked fourth, with unbanned renewable energy and residential land use ranked at second and third randomly. If both solar and wind energy projects are banned in the county, residential land use is ranked second, with solar and wind ranked third and fourth randomly. If renewable energy projects are permitted, residential use is ranked fourth, and wind and solar are ranked second or third based on restriction levels. If restrictions are equal, the ranking is assigned randomly.

\begin{figure}[htbp]
\begin{centering}
\includegraphics[scale=.45]{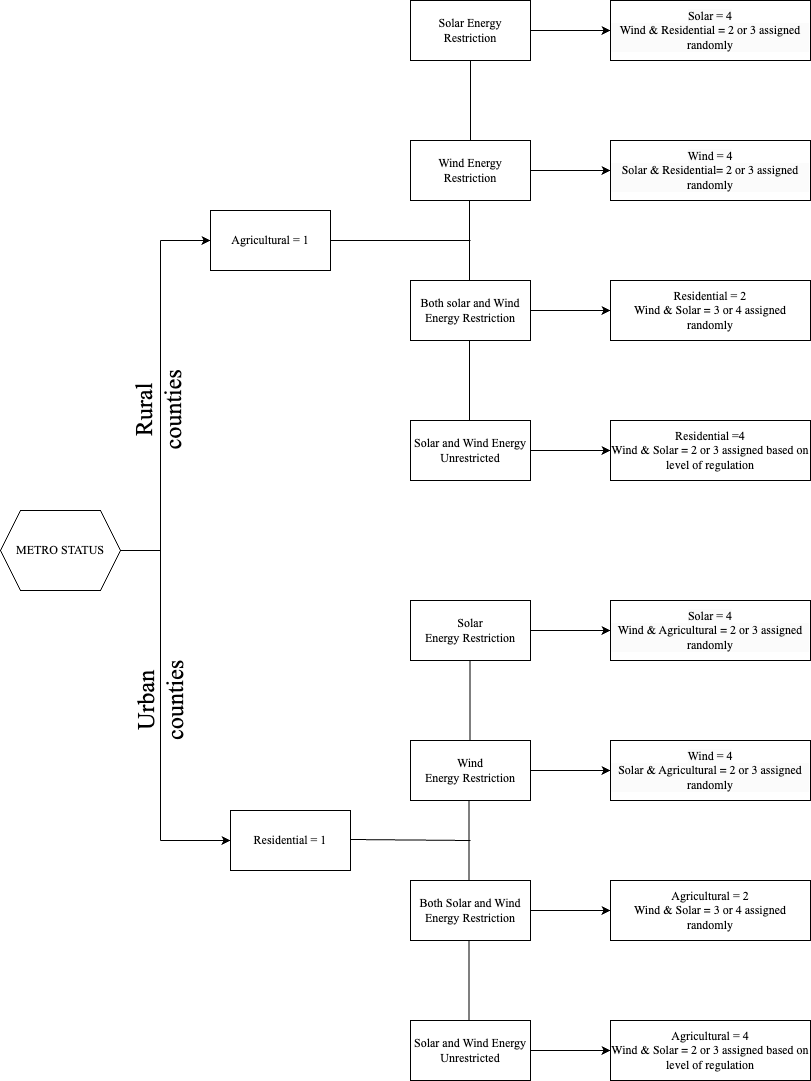}
\caption{Flowchart showing the ranking of land use preference}
\label{flowchart}
\end{centering}
\end{figure}

In urban counties, residential land use is ranked first. We then examine renewable energy regulations to rank agricultural, solar, and wind land use. If either solar or wind energy is banned, the banned category is ranked fourth, with unbanned renewable and agricultural land use ranked at second and third randomly. If both solar and wind are banned, agricultural land use is ranked second, with solar and wind ranked third and fourth randomly. If renewable energy projects are permitted, agricultural use is ranked fourth, and wind and solar are ranked second or third based on restriction levels. If restrictions are equal, the ranking is assigned randomly. The ranking procedure is illustrated in figure \ref{flowchart}.

Table \ref{land_use_ranking} reports the land use preferences and the number of observations for each land use type at specified ranking during our study period. Agriculture ranks in the first position for 2550 observations. It also appears in the second, third, and fourth positions for 475, 479, and 454 observations, respectively. Solar and Wind energy do not occupy any first rankings. However, Solar energy ranks second, third, and fourth for 1348, 1286, and 1324 observations, respectively. Similarly, Wind energy ranks second, third, and fourth for 1272, 1344, and 1342 observations, respectively. Residential land use ranks first in 1408 observations and ranks second, third, and fourth for 863, 849, and 838 observations, respectively.

\begin{table}
\centering
  \caption{Land Use Preference Ranking}
  \label{land_use_ranking}
{
\begin{tabular}{lcccc}
\\[-1.8ex]\hline 
\hline \\[-1.8ex] 
\textit{Land Use}  & \multicolumn{3}{c}{\textit{Ranking:}} & \\ 
\\[-1.8ex] & 1 & 2 & 3 & 4 \\ 
\hline \\[-1.8ex] 
Agriculture & 2550 & 475 & 479 & 454 \\ 
& & & &  \\
Solar Energy & 0 & 1348 & 1286 & 1324 \\ 
& & & &  \\
Wind Energy & 0 & 1272 & 1344 & 1342 \\ 
& & & &  \\
Residential & 1408 & 863 & 849 & 838 \\ 
\hline 
\hline \\[-1.8ex] 
\end{tabular}}
\end{table}

\subsection{Estimation Results}

Table \ref{tab:first_stage} reports the ordinary least squares estimates for the relationship between number of days with high temperatures and returns to corn farming in each county. These estimates are the first stage in our two stage lease squares estimates of the cross return elasticity of farming and acceptance of wind and solar leases. Column one reports the estimates for the described relationship without taking into account the preferences of the landowner. The estimated relationship is negative and statistically different than zero, satisfying \textbf{Assumption I3} (full rank or relevance). However, as discussed in our identification section, it is important to examine the role of preferences to estimate economically meaningful parameters. As a result, columns two, three, four, and five report the OLS estimates of the relationship between number of days with high temperatures and returns to corn farming in each county, conditional on the second best option being Agriculture, Residential, Solar, and Wind, respectively. The estimate for each second-best option is negative and of similar magnitude, but the estimate for Residential as the second-best option is only weakly significant. 

\begin{table}[htbp]\centering
\def\sym#1{\ifmmode^{#1}\else\(^{#1}\)\fi}
\caption{First Stage estimates of High Heat Days as an instrument for Return to Farming}\label{tab:first_stage}
\begin{tabular}{l*{5}{c}}
\toprule
                    &       & \multicolumn{4}{c}{Land Use Ranked Second}  \\

                    &\multicolumn{1}{c}{ALL}         &\multicolumn{1}{c}{Agriculture}         &\multicolumn{1}{c}{Residential}         &\multicolumn{1}{c}{Solar}         &\multicolumn{1}{c}{Wind}         \\
\midrule
High Heat Days      &      -0.006\sym{***}&      -0.010\sym{***}&      -0.003\sym{*}  &      -0.007\sym{***}&      -0.004\sym{***}\\
                    &     (0.001)         &     (0.002)         &     (0.002)         &     (0.001)         &     (0.001)         \\
\addlinespace
Median Census Age   &       0.002         &      -0.005         &       0.008         &      -0.010         &       0.007         \\
                    &     (0.004)         &     (0.018)         &     (0.009)         &     (0.009)         &     (0.008)         \\
\addlinespace
Unemployment Rate   &       0.004         &      -0.001         &       0.009         &      -0.004         &       0.001         \\
                    &     (0.003)         &     (0.012)         &     (0.007)         &     (0.006)         &     (0.007)         \\
\addlinespace
Constant            &       6.324\sym{***}&       6.628\sym{***}&       6.030\sym{***}&       6.935\sym{***}&       6.115\sym{***}\\
                    &     (0.167)         &     (0.842)         &     (0.360)         &     (0.370)         &     (0.336)         \\
\midrule
Obs.                &    3958.000         &     475.000         &     863.000         &    1348.000         &    1272.000         \\
County Fixed Effects&         yes         &         yes         &         yes         &         yes         &         yes         \\
Year Fixed Effects  &         yes         &         yes         &         yes         &         yes         &         yes         \\
$R^2$                 &       0.907         &       0.935         &       0.945         &       0.920         &       0.930         \\
\bottomrule
\multicolumn{6}{l}{\footnotesize Standard errors in parentheses}\\
\multicolumn{6}{l}{\footnotesize \sym{*} \(p<0.10\), \sym{**} \(p<0.05\), \sym{***} \(p<0.010\)}\\
\end{tabular}
\end{table}

The estimates of the first stage reported in table \ref{tab:first_stage} highlight the potential importance of conditioning our estimation on landowners' preferences. Although, the estimated coefficients are of similar sign, they are of varying magnitudes. For example, in counties where agriculture is the second best option for land use, the instrument is shown to have a larger negative relationship with our dependent variable, relative to counties where residential, solar, and wind are the secondary preference. 

Table \ref{tab:extensive_solar_w_rank_2} reports the two-stage least squares estimates of equation (\ref{eq:choi_str}) for the cross-return elasticity of farming and accepting solar leases.  In a similar convention to table \ref{tab:first_stage}, the first column reports the estimates without considering landowners' preference ranking, and columns two, three, four, and five report the estimates taking into account landowners' preferences. When estimated without preferences, it does not appear that landowners' decision to accept solar leases are not influenced by short-term returns to farming, or the landowners' cross-return elasticity for returns to farming and solar energy transition is not statistically different from zero. In contrast, the cross-return elasticity for landowners in counties where agriculture is the second best option (column two) is negative and in a magnitude that we would expect. A one percent increase in the returns to corn farming decreases the likelihood that a landowner will accept solar leases by one percent. Based upon our ranking procedure, these landowners' are in residential counties with regulations that make developing wind and solar energy difficult. Landowners' willingness to accept solar leases in the remaining alternatives are not influenced by returns to farming.

\begin{table}[htbp]\centering
\def\sym#1{\ifmmode^{#1}\else\(^{#1}\)\fi}
\caption{Cross Return Elasticity of Farming and Solar Adoption}\label{tab:extensive_solar_w_rank_2}
\begin{tabular}{l*{5}{c}}
\toprule
                    &       & \multicolumn{4}{c}{Land Use Ranked Second}  \\

                    &\multicolumn{1}{c}{ALL}         &\multicolumn{1}{c}{Agriculture}         &\multicolumn{1}{c}{Residential}         &\multicolumn{1}{c}{Solar}         &\multicolumn{1}{c}{Wind}         \\
\midrule
Farm Return         &      -0.132         &      -1.006\sym{**} &      -0.680         &       0.448         &      -0.636         \\
                    &     (0.184)         &     (0.404)         &     (0.897)         &     (0.299)         &     (0.511)         \\
\addlinespace
Median Census Age   &       0.032\sym{***}&       0.095\sym{***}&       0.032\sym{*}  &       0.019         &       0.029\sym{**} \\
                    &     (0.007)         &     (0.032)         &     (0.016)         &     (0.013)         &     (0.014)         \\
\addlinespace
Unemployment Rate   &       0.034\sym{***}&       0.094\sym{***}&       0.037\sym{**} &       0.038\sym{***}&       0.018\sym{*}  \\
                    &     (0.005)         &     (0.022)         &     (0.014)         &     (0.009)         &     (0.011)         \\
\addlinespace
Constant            &      -0.718         &       3.502         &       2.761         &      -3.919\sym{*}  &       2.636         \\
                    &     (1.183)         &     (2.871)         &     (5.352)         &     (2.113)         &     (3.140)         \\
\midrule
Obs.                &    3958.000         &     475.000         &     863.000         &    1348.000         &    1272.000         \\
County Fixed Effects&         yes         &         yes         &         yes         &         yes         &         yes         \\
Year Fixed Effects  &         yes         &         yes         &         yes         &         yes         &         yes         \\
$R^2$                 &       0.717         &       0.754         &       0.765         &       0.777         &       0.761         \\
\bottomrule
\multicolumn{6}{l}{\footnotesize Standard errors in parentheses}\\
\multicolumn{6}{l}{\footnotesize \sym{*} \(p<0.10\), \sym{**} \(p<0.05\), \sym{***} \(p<0.010\)}\\
\end{tabular}
\end{table}

We believe these results are driven by the long-term potential returns to residential or commercial development. If landowners' are assumed to view the value of their farmland as the cumulative future value, then the estimates of column two suggest that landowners' long-term view of the option for future development with the added returns to farming are greater than the offered returns through solar leases. Hence, solar leases payments are large enough to equate to the long-term potential of residential or commercial development, or, for these landowners solar leases are at the monetary decision margin.

Table \ref{tab:extensive_wind_w_rank_2} reports the two-stage least squares estimates of equation (\ref{eq:choi_str}) for the cross return elasticity of returns to farming and acceptance of wind leases. Table \ref{tab:extensive_wind_w_rank_2} follows the same convention of tables \ref{tab:first_stage} and \ref{tab:extensive_solar_w_rank_2}, estimates reported in column one do not account for landowner preferences, and columns two, three, four, and five report estimates corresponding to counties where the specified option is second best. In contrast to the estimates of cross-return elasticity of returns to farming and solar, when preferences are not taken into account, the estimates of column one suggest that an increase in returns to farming corn is negatively related to a landowner accepting wind energy leases, but with weak significance. We believe this estimate suggests that landowners' willingness to accept wind leases, at the current pricing, is nearer to the decision margin. In contrast to solar leases, farmland still retains some capacity for farming, but with the added encumbrance of farming around turbines and access roads. Hence, a short-term increase in returns to farming, can be perceived as greater than the long-term combination of returns to farming in combination with wind lease. 

\begin{table}[htbp]\centering
\def\sym#1{\ifmmode^{#1}\else\(^{#1}\)\fi}
\caption{Cross Return Elasticity of Farming and Wind Adoption}\label{tab:extensive_wind_w_rank_2}
\begin{tabular}{l*{5}{c}}
\toprule
                    &       & \multicolumn{4}{c}{Land Use Ranked Second}  \\

                    &\multicolumn{1}{c}{ALL}         &\multicolumn{1}{c}{Agriculture}         &\multicolumn{1}{c}{Residential}         &\multicolumn{1}{c}{Solar}         &\multicolumn{1}{c}{Wind}         \\
\midrule
Farm Return         &      -0.286\sym{*}  &      -0.853\sym{***}&       2.139         &      -0.747\sym{**} &      -0.348         \\
                    &     (0.168)         &     (0.235)         &     (1.375)         &     (0.296)         &     (0.427)         \\
\addlinespace
Median Census Age   &      -0.022\sym{***}&       0.006         &      -0.031         &      -0.017         &      -0.050\sym{***}\\
                    &     (0.006)         &     (0.019)         &     (0.025)         &     (0.013)         &     (0.012)         \\
\addlinespace
Unemployment Rate   &       0.010\sym{**} &       0.024\sym{*}  &       0.004         &       0.004         &       0.005         \\
                    &     (0.005)         &     (0.013)         &     (0.022)         &     (0.009)         &     (0.009)         \\
\addlinespace
Constant            &       2.621\sym{**} &       5.080\sym{***}&     -12.560         &       5.511\sym{***}&       4.148         \\
                    &     (1.078)         &     (1.672)         &     (8.202)         &     (2.095)         &     (2.625)         \\
\midrule
Obs.                &    3958.000         &     475.000         &     863.000         &    1348.000         &    1272.000         \\
County Fixed Effects&         yes         &         yes         &         yes         &         yes         &         yes         \\
Year Fixed Effects  &         yes         &         yes         &         yes         &         yes         &         yes         \\
$R^2$                 &       0.728         &       0.844         &       0.555         &       0.744         &       0.792         \\
\bottomrule
\multicolumn{6}{l}{\footnotesize Standard errors in parentheses}\\
\multicolumn{6}{l}{\footnotesize \sym{*} \(p<0.10\), \sym{**} \(p<0.05\), \sym{***} \(p<0.010\)}\\
\end{tabular}
\end{table}

In a similar pattern to the estimates of table \ref{tab:extensive_solar_w_rank_2}, landowners' willingness to accept wind leases in residential counties with wind and solar regulations is influenced by short-term returns to farming. A one percent increase in returns to farming corn decreases the likelihood that a landowner accepts a wind lease by 0.853 percent. We believe this results is driven by the long-term potential to develop the farmland for residential or commercial purposes. As such, these landowners' decisions appear to be at the monetary margin. 

In contrast to the estimates of table \ref{tab:extensive_solar_w_rank_2}, landowners' willingness to accept wind leases in counties where solar energy is the second best option (column four) farm returns have a negative relationship with acceptance. A one percent increase in returns to farming is estimated to decrease the likelihood of accepting wind leases by 0.747 percent. Based on our ranking procedure, these landowners could be from one of two scenarios. In scenario one, a landowner is in a residential county with wind regulations. In this scenario, we believe the results are driven by similar logic as discussed previously, residential and commercial development potential. In scenario two, a landowner is in an agricultural county, with wind regulations. In this scenario, we must assume that a short-term increase in returns to farming, are large enough to outweigh the long-term returns to farming in combination with wind lease return. Perhaps, the long-term encumbrance of farming around turbines and access roads. 

The estimates in column one of both tables \ref{tab:extensive_solar_w_rank_2} and \ref{tab:extensive_wind_w_rank_2}, suggest that wind leases are nearer the landowners' monetary decision margin. We take this result, as suggestive evidence that landowners' are taking into account the degree to which the option to still engage in production agriculture is present. As highlighted previously, wind energy generation still allows the landowner to generate revenue from farming, whereas solar energy generation removes this potential for the life of the lease. This optionality of wind energy generation is an important factor in the monetary decision margin.

The results of tables \ref{tab:first_stage}, \ref{tab:extensive_solar_w_rank_2}, and \ref{tab:extensive_wind_w_rank_2} illustrate the importance of considering the preferences of landowners, specifically when the set of options is larger than two. As discussed previously, with regard to land conversion to solar energy generation, the cross return elasticity for farming is zero when not taking into account landowners' preferences. However, our estimates suggest that landowners' willingness to accept solar leases are sensitive to changes in returns to farming in residential counties with less favorable wind and solar regulations. With regard to land conversions to wind energy generation, the cross return elasticity for farming is negative and weakly significant when not taking into account landowners' preferences. When taking into account landowners' preferences, our estimates suggest that landowners' willingness to accept wind leases are influenced by returns to farming in residential counties with less favorable wind and solar regulations and in agricultural counties with less favorable wind regulations. Thus, we believe that our proposed approach more accurately identifies the relationship between returns to farming and acceptance of leases for wind and solar.  

We perform estimate equation (\ref{eq:choi_str}) with several different variations to ensure that our results are not driven by our modeling choices. First, we vary our instrumental variable, by changing the threshold for \textit{High Heat Days}. In our primary specification we use 83 degrees, consistent with \citet{schlenker2009nonlinear}. This modeling choice is not driving our results. For the sake of space, the results are included in the appendix. Second, we generate a new ranking of second-best option, by changing the randomization. Similar to the choice of heat threshold, generating a new randomized ranking of second-best land use option, suggests this modeling choice is not driving our results.  

\section{Conclusion} \label{sec:cls}

This study proposes a new econometric method to estimate treatment effects accounting for an economic agent's outside options. We use this empirical approach to examine a key land-use issue, the use of farmland for renewable energy generation, namely wind and solar. With a unique dataset, we show that landowners' willingness to accept wind and solar leases are only sensitive to short-term returns to farming in limited scenarios.This analysis underscores the importance of considering agents' options and preferences when modeling decision-making. Indeed, neglecting agents' preferences can lead to an overestimation of the significance of monetary factors in instrumental variable (IV) estimates. In contrast, our estimation—guided by the recommendations outlined in our identification section—reveals a more nuanced perspective.


This study uses a short panel of county-level aggregate utility scale solar and wind installations. As the number of utility scale solar and wind installations grows, future research can exploit this additional time series. Additionally, using surveys or working with wind and solar developers to understand actual contracts offered to landowners would help to accurately identify the non-monetary dimensions on which landowners derive value.  
 
We believe this research has significant implications for land-use policy. Land use is a critical issue, as policymakers must balance environmental concerns, food security, and the needs of private citizens. Our findings suggest that, the effectiveness of policy interventions will be directly related to landowners' alternative land-uses. For urban counties, if the desire is to increase the development of solar or wind energy generation, policymakers can look toward tax credits or subsidies, effectively increasing the returns to farmland owners. For rural counties dominated by agricultural land, the policy scenario is more complex. As mentioned earlier, the landowners' decision are not at the monetary margin. Hence, if the county is looking to increase or decrease solar or wind energy development, policymakers will need to use additional zoning regulations. In the case of a desired increase in solar or wind development, policymakers should consider mechanisms which may account for the non-monetary benefits that landowners' may receive for their farmland remaining in agricultural production. Additionally, in counties with societal opposition to renewable energy development, policymakers may consider schemes which benefit all citizens, as opposed to solely the private landowners.

\newpage
\bibliographystyle{chicago}

\bibliography{citations}

\newpage
\section*{Appendix}

\begin{table}[htbp]\centering
\def\sym#1{\ifmmode^{#1}\else\(^{#1}\)\fi}
\caption{Cross Return Elasticity of Farming and Solar Adoption with Days above 80}
\begin{tabular}{l*{5}{c}}
\toprule
                    &       & \multicolumn{4}{c}{Land Use Ranked Second}  \\

                    &\multicolumn{1}{c}{ALL}         &\multicolumn{1}{c}{Agriculture}         &\multicolumn{1}{c}{Residential}         &\multicolumn{1}{c}{Solar}         &\multicolumn{1}{c}{Wind}         \\
\midrule
Farm Return         &      -0.156         &      -0.756\sym{*}  &      -0.297         &      -0.017         &      -0.565         \\
                    &     (0.159)         &     (0.419)         &     (0.682)         &     (0.203)         &     (0.467)         \\
\addlinespace
Median Census Age   &       0.032\sym{***}&       0.096\sym{***}&       0.028\sym{*}  &       0.015         &       0.029\sym{**} \\
                    &     (0.007)         &     (0.031)         &     (0.015)         &     (0.013)         &     (0.014)         \\
\addlinespace
Unemployment Rate   &       0.034\sym{***}&       0.094\sym{***}&       0.033\sym{***}&       0.037\sym{***}&       0.018\sym{*}  \\
                    &     (0.005)         &     (0.021)         &     (0.013)         &     (0.009)         &     (0.011)         \\
\addlinespace
Constant            &      -0.569         &       1.872         &       0.486         &      -0.746         &       2.201         \\
                    &     (1.029)         &     (2.945)         &     (4.078)         &     (1.486)         &     (2.881)         \\
\midrule
Obs.                &    3958.000         &     475.000         &     863.000         &    1348.000         &    1272.000         \\
County Fixed Effects&         yes         &         yes         &         yes         &         yes         &         yes         \\
Year Fixed Effects  &                     &                     &                     &                     &                     \\
$R^2$                 &       0.716         &       0.777         &       0.791         &       0.790         &       0.766         \\
\bottomrule
\multicolumn{6}{l}{\footnotesize Standard errors in parentheses}\\
\multicolumn{6}{l}{\footnotesize \sym{*} \(p<0.10\), \sym{**} \(p<0.05\), \sym{***} \(p<0.010\)}\\
\end{tabular}
\end{table}

\begin{table}[htbp]\centering
\def\sym#1{\ifmmode^{#1}\else\(^{#1}\)\fi}
\caption{Cross Return Elasticity of Farming and Solar Adoption with days Above 86}
\begin{tabular}{l*{5}{c}}
\toprule
                    &       & \multicolumn{4}{c}{Land Use Ranked Second}  \\

                    &\multicolumn{1}{c}{ALL}         &\multicolumn{1}{c}{Agriculture}         &\multicolumn{1}{c}{Residential}         &\multicolumn{1}{c}{Solar}         &\multicolumn{1}{c}{Wind}         \\
\midrule
Farm Return         &      -0.387         &      -1.150\sym{**} &      -1.275         &       3.210         &       1.752         \\
                    &     (0.727)         &     (0.447)         &     (4.992)         &     (4.527)         &     (2.700)         \\
\addlinespace
Median Census Age   &       0.033\sym{***}&       0.094\sym{***}&       0.038         &       0.041         &       0.010         \\
                    &     (0.008)         &     (0.033)         &     (0.054)         &     (0.044)         &     (0.028)         \\
\addlinespace
Unemployment Rate   &       0.035\sym{***}&       0.094\sym{***}&       0.043         &       0.043\sym{**} &       0.018         \\
                    &     (0.007)         &     (0.023)         &     (0.052)         &     (0.021)         &     (0.014)         \\
\addlinespace
Constant            &       0.867         &       4.446         &       6.287         &     -22.779         &     -11.831         \\
                    &     (4.522)         &     (3.147)         &    (29.606)         &    (30.932)         &    (16.371)         \\
\midrule
Obs.                &    3958.000         &     475.000         &     863.000         &    1348.000         &    1272.000         \\
County Fixed Effects&         yes         &         yes         &         yes         &         yes         &         yes         \\
Year Fixed Effects  &                     &                     &                     &                     &                     \\
$R^2$                 &       0.704         &       0.738         &       0.690         &       0.109         &       0.615         \\
\bottomrule
\multicolumn{6}{l}{\footnotesize Standard errors in parentheses}\\
\multicolumn{6}{l}{\footnotesize \sym{*} \(p<0.10\), \sym{**} \(p<0.05\), \sym{***} \(p<0.010\)}\\
\end{tabular}
\end{table}

\begin{table}[htbp]\centering
\def\sym#1{\ifmmode^{#1}\else\(^{#1}\)\fi}
\caption{Cross Return Elasticity of Farming and Solar Adoption with Alternative Ranking}
\begin{tabular}{l*{5}{c}}
\toprule
                    &       & \multicolumn{4}{c}{Land Use Ranked Second}  \\

                    &\multicolumn{1}{c}{ALL}         &\multicolumn{1}{c}{Agriculture}         &\multicolumn{1}{c}{Residential}         &\multicolumn{1}{c}{Solar}         &\multicolumn{1}{c}{Wind}         \\
\midrule
Farm Return         &      -0.132         &      -0.771\sym{**} &      -0.423         &      -0.037         &       0.411\sym{**} \\
                    &     (0.184)         &     (0.380)         &     (0.994)         &     (0.409)         &     (0.199)         \\
\addlinespace
Median Census Age   &       0.032\sym{***}&       0.024         &      -0.004         &       0.039\sym{***}&       0.040\sym{***}\\
                    &     (0.007)         &     (0.034)         &     (0.014)         &     (0.014)         &     (0.011)         \\
\addlinespace
Unemployment Rate   &       0.034\sym{***}&       0.022         &       0.025\sym{*}  &       0.050\sym{***}&       0.017\sym{*}  \\
                    &     (0.005)         &     (0.023)         &     (0.013)         &     (0.010)         &     (0.009)         \\
\addlinespace
Constant            &      -0.718         &       4.792\sym{*}  &       2.644         &      -1.741         &      -4.541\sym{***}\\
                    &     (1.183)         &     (2.767)         &     (6.443)         &     (2.465)         &     (1.275)         \\
\midrule
Obs.                &    3958.000         &     464.000         &     855.000         &    1309.000         &    1305.000         \\
County Fixed Effects&         yes         &         yes         &         yes         &         yes         &         yes         \\
Year Fixed Effects  &                     &                     &                     &                     &                     \\
$R^2$                 &       0.717         &       0.793         &       0.731         &       0.792         &       0.790         \\
\bottomrule
\multicolumn{6}{l}{\footnotesize Standard errors in parentheses}\\
\multicolumn{6}{l}{\footnotesize \sym{*} \(p<0.10\), \sym{**} \(p<0.05\), \sym{***} \(p<0.010\)}\\
\end{tabular}
\end{table}

\begin{table}[htbp]\centering
\def\sym#1{\ifmmode^{#1}\else\(^{#1}\)\fi}
\caption{Cross Return Elasticity of Farming and Wind Adoption with Days above 80}
\begin{tabular}{l*{5}{c}}
\toprule
                    &       & \multicolumn{4}{c}{Land Use Ranked Second}  \\

                    &\multicolumn{1}{c}{ALL}         &\multicolumn{1}{c}{Agriculture}         &\multicolumn{1}{c}{Residential}         &\multicolumn{1}{c}{Solar}         &\multicolumn{1}{c}{Wind}         \\
\midrule
Farm Return         &      -0.476\sym{***}&      -0.647\sym{***}&      -0.721         &      -0.371\sym{*}  &      -0.953\sym{**} \\
                    &     (0.147)         &     (0.232)         &     (0.793)         &     (0.195)         &     (0.440)         \\
\addlinespace
Median Census Age   &      -0.022\sym{***}&       0.007         &      -0.002         &      -0.014         &      -0.045\sym{***}\\
                    &     (0.007)         &     (0.017)         &     (0.017)         &     (0.012)         &     (0.013)         \\
\addlinespace
Unemployment Rate   &       0.011\sym{**} &       0.024\sym{**} &       0.033\sym{**} &       0.005         &       0.005         \\
                    &     (0.005)         &     (0.012)         &     (0.015)         &     (0.009)         &     (0.010)         \\
\addlinespace
Constant            &       3.804\sym{***}&       3.733\sym{**} &       4.398         &       2.939\sym{**} &       7.814\sym{***}\\
                    &     (0.955)         &     (1.632)         &     (4.746)         &     (1.430)         &     (2.712)         \\
\midrule
Obs.                &    3958.000         &     475.000         &     863.000         &    1348.000         &    1272.000         \\
County Fixed Effects&         yes         &         yes         &         yes         &         yes         &         yes         \\
Year Fixed Effects  &                     &                     &                     &                     &                     \\
$R^2$                 &       0.717         &       0.872         &       0.772         &       0.773         &       0.741         \\
\bottomrule
\multicolumn{6}{l}{\footnotesize Standard errors in parentheses}\\
\multicolumn{6}{l}{\footnotesize \sym{*} \(p<0.10\), \sym{**} \(p<0.05\), \sym{***} \(p<0.010\)}\\
\end{tabular}
\end{table}

\begin{table}[htbp]\centering
\def\sym#1{\ifmmode^{#1}\else\(^{#1}\)\fi}
\caption{Cross Return Elasticity of Farming and Wind Adoption with days Above 86}
\begin{tabular}{l*{5}{c}}
\toprule
                    &       & \multicolumn{4}{c}{Land Use Ranked Second}  \\

                    &\multicolumn{1}{c}{ALL}         &\multicolumn{1}{c}{Agriculture}         &\multicolumn{1}{c}{Residential}         &\multicolumn{1}{c}{Solar}         &\multicolumn{1}{c}{Wind}         \\
\midrule
Farm Return         &      -0.319         &      -1.072\sym{***}&      -6.075         &      -1.287         &      -1.578         \\
                    &     (0.650)         &     (0.281)         &    (13.864)         &     (2.573)         &     (2.353)         \\
\addlinespace
Median Census Age   &      -0.022\sym{***}&       0.005         &       0.054         &      -0.021         &      -0.040\sym{*}  \\
                    &     (0.007)         &     (0.021)         &     (0.150)         &     (0.025)         &     (0.024)         \\
\addlinespace
Unemployment Rate   &       0.010\sym{*}  &       0.024\sym{*}  &       0.088         &       0.003         &       0.005         \\
                    &     (0.006)         &     (0.014)         &     (0.146)         &     (0.012)         &     (0.012)         \\
\addlinespace
Constant            &       2.825         &       6.513\sym{***}&      36.150         &       9.199         &      11.599         \\
                    &     (4.045)         &     (1.982)         &    (82.229)         &    (17.582)         &    (14.265)         \\
\midrule
Obs.                &    3958.000         &     475.000         &     863.000         &    1348.000         &    1272.000         \\
County Fixed Effects&         yes         &         yes         &         yes         &         yes         &         yes         \\
Year Fixed Effects  &                     &                     &                     &                     &                     \\
$R^2$                 &       0.727         &       0.805         &           .         &       0.664         &       0.635         \\
\bottomrule
\multicolumn{6}{l}{\footnotesize Standard errors in parentheses}\\
\multicolumn{6}{l}{\footnotesize \sym{*} \(p<0.10\), \sym{**} \(p<0.05\), \sym{***} \(p<0.010\)}\\
\end{tabular}
\end{table}

\begin{table}[htbp]\centering
\def\sym#1{\ifmmode^{#1}\else\(^{#1}\)\fi}
\caption{Cross Return Elasticity of Farming and Wind Adoption with Alternative Ranking}
\begin{tabular}{l*{5}{c}}
\toprule
                    &       & \multicolumn{4}{c}{Land Use Ranked Second}  \\

                    &\multicolumn{1}{c}{ALL}         &\multicolumn{1}{c}{Agriculture}         &\multicolumn{1}{c}{Residential}         &\multicolumn{1}{c}{Solar}         &\multicolumn{1}{c}{Wind}         \\
\midrule
Farm Return         &      -0.286\sym{*}  &      -0.890\sym{***}&      -0.170         &       0.045         &      -0.019         \\
                    &     (0.168)         &     (0.221)         &     (0.968)         &     (0.347)         &     (0.198)         \\
\addlinespace
Median Census Age   &      -0.022\sym{***}&       0.030         &      -0.026\sym{*}  &      -0.053\sym{***}&      -0.037\sym{***}\\
                    &     (0.006)         &     (0.020)         &     (0.014)         &     (0.012)         &     (0.011)         \\
\addlinespace
Unemployment Rate   &       0.010\sym{**} &       0.004         &       0.022\sym{*}  &       0.006         &       0.012         \\
                    &     (0.005)         &     (0.013)         &     (0.013)         &     (0.009)         &     (0.009)         \\
\addlinespace
Constant            &       2.621\sym{**} &       4.555\sym{***}&       1.935         &       1.832         &       1.419         \\
                    &     (1.078)         &     (1.607)         &     (6.273)         &     (2.091)         &     (1.269)         \\
\midrule
Obs.                &    3958.000         &     464.000         &     855.000         &    1309.000         &    1305.000         \\
County Fixed Effects&         yes         &         yes         &         yes         &         yes         &         yes         \\
Year Fixed Effects  &                     &                     &                     &                     &                     \\
$R^2$                 &       0.728         &       0.805         &       0.799         &       0.830         &       0.789         \\
\bottomrule
\multicolumn{6}{l}{\footnotesize Standard errors in parentheses}\\
\multicolumn{6}{l}{\footnotesize \sym{*} \(p<0.10\), \sym{**} \(p<0.05\), \sym{***} \(p<0.010\)}\\
\end{tabular}
\end{table}

\end{document}